\documentclass[pra,twocolumn,showpacs,superscriptaddress]{revtex4}

\usepackage{amsmath}
\usepackage{amsfonts}
\usepackage{amssymb}
\usepackage[dvips]{graphicx}

\begin{document}

\title{Mutually unbiased bases, orthogonal Latin squares, and hidden-variable models}

\author{Tomasz Paterek}
\affiliation{Institute for Quantum Optics and Quantum Information,
Austrian Academy of Sciences, Boltzmanngasse 3, A-1090 Vienna,
Austria}

\author{Borivoje Daki\'{c}}
\affiliation{Institute for Quantum Optics and Quantum Information,
Austrian Academy of Sciences, Boltzmanngasse 3, A-1090 Vienna,
Austria} \affiliation{Faculty of Physics, University of Vienna,
Boltzmanngasse 5, A-1090 Vienna, Austria}

\author{{\v C}aslav Brukner}
\affiliation{Institute for Quantum Optics and Quantum Information,
Austrian Academy of Sciences, Boltzmanngasse 3, A-1090 Vienna,
Austria} \affiliation{Faculty of Physics, University of Vienna,
Boltzmanngasse 5, A-1090 Vienna, Austria}

\date{\today}

\begin{abstract}
Mutually unbiased bases encapsulate the concept of complementarity 
--- the impossibility of simultaneous knowledge of certain observables --- 
in the formalism of quantum theory.
Although this concept is at the heart of quantum mechanics, the number of these bases is unknown
except for systems of dimension being a power of a prime.
We develop the relation between this physical problem and the mathematical problem of finding the number of mutually orthogonal Latin squares.
We derive in a simple way all known results about the unbiased bases, 
find their lower number, 
and disprove the existence of certain forms of the bases in dimensions different than power of a prime.
Using the Latin squares, we construct hidden-variable models
which efficiently simulate results of complementary quantum measurements.
\end{abstract}


\maketitle

\section{Introduction}

Complementarity is a fundamental principle of quantum physics
which forbids simultaneous knowledge of certain observables.
It is manifested already for the simplest
quantum mechanical system --- spin-$\frac{1}{2}$.
If the system is in a definite state of, say, spin along $x$,
the spin along $y$ or $z$ is completely unknown,
i.e.,\ the outcomes ``spin up'' and ``spin down''
occur with the same probability.
The eigenbases of $\hat \sigma_x$, $\hat \sigma_y$ and $\hat \sigma_z$
Pauli operators form so-called mutually unbiased bases (MUBs):
Every vector from one basis has equal overlap with all the vectors from other bases.
MUBs encapsulate the concept of complementarity in the quantum formalism.
Although complementarity is at the heart of quantum physics,
the question about the number of MUBs remains unanswered.
Apart from being of foundational interest, 
MUBs find applications in quantum state tomography \cite{WOOTTERSFIELDS},
quantum-key distribution \cite{QKD_REVIEW}, and the mean King problem \cite{MEAN_KING}.

A $d$-level quantum system can have at most $d+1$ MUBs,
and such a set is referred to as the complete set of MUBs.
In 1981 Ivanovi{\' c} proved by construction that there are indeed $d+1$ complementary measurements for $d$ being a prime number \cite{IVANOVIC}. 
This result was generalized by Wootters and Fields to cover powers of primes \cite{WOOTTERSFIELDS}.
For other dimensions the number of MUBs is unknown,
the simplest case being dimension six.
A considerable amount of work was done towards understanding this problem.
New proofs of previous results were established \cite{UNITARY_MUB,CASLAVS,KR,DURT} 
and the problem was linked with other unsolved problems \cite{LIE,HADAMARD}.
It was also noticed that it is similar in spirit to certain problems in combinatorics \cite{ZAUNER,BENGTSSON,WB} 
and finite geometry \cite{FINITE_GEOM}.
Here, we build upon these relations.

We describe the problem of the number of orthogonal Latin squares (OLSs),
which was initiated by Euler \cite{EULER} and still attracts lots of attention in mathematics.
Although this problem is not solved yet in full generality,
more is known about it than about the number of MUBs.
Using a black box which physically encodes
information contained in a Latin square,
we link every OLS of order being a power of a prime with a MUB.
For dimension six, our method gives three MUBs,
which is the maximal number found by the numerical research \cite{ZAUNER,HADAMARD}.
Utilizing known results for OLSs we derive a minimal number of MUBs,
and disprove the existence of certain forms of MUBs for arbitrary $d$.
Finally, using OLSs we construct hidden-variable models
that efficiently simulate complementary quantum measurements.

\section{Orthogonal Latin squares}

A Latin square of order $d$ is an array of numbers $\{0,...,d-1\}$
where every row and every column contains each number exactly once.
Two Latin squares, $A=[A_{ij}]$ and $B=[B_{ij}]$, are orthogonal 
if all \emph{ordered} pairs $(A_{ij},B_{ij})$ are distinct.
There are at most $d-1$ OLSs and this set is called complete.
The existence of $L$ OLSs is equivalent to the existence
of a combinatorial design called a \emph{net} with $L+2$ rows \cite{DESIGN}.
The net design has a form of a table 
in which every row contains $d^2$ distinct numbers.
They are split into $d$ cells of $d$ numbers each,
in such a way that the numbers of any cell in a given row
are distributed among all cells of any other row.
The additional two rows of the net
correspond to orthogonal but not Latin squares,
with the entries $A_{ij} = j$ and $A_{ij} = i$.

The following algorithm allows us to construct the net from a set of OLSs:

(i) Write the squares in the standard form in which 
the numbers of the first column are in ascending order
(by permuting the entries, it is always possible to write the set of OLSs in the standard form without compromising Latiness and orthogonality).

(ii) Augment the set of OLSs by the two orthogonal non-Latin squares $A_{ij} = j$ and $A_{ij} = i$.

(iii) Write the rows of the squares as cells in a single row of the table. 
The number of the table's rows is now equal to the number of squares in the augmented set.

(iv) In the row of the table which corresponds to the square $A_{ij} = j$, referred to as the ``coordinate row,''
replace the number $A_{ij}$ in the $i$th cell by $A_{ij}' = i d+j$, where $d$ is the order of the square.

(v) In every cell of the other rows replace number $B_{ij}$
on position $j$ by the integer associated to the number $B_{ij}$
of the $j$th cell in the coordinate row, i.e., $B_{ij} \to B_{ij}' = j d + B_{ij}$.

We shall prove that the table generated by this procedure is indeed a net design.
We use another property defining the design:
Two numbers in one cell do not repeat in any other cell.
This already includes that any two cells of two different rows
share exactly one common number, as if there were no common numbers shared
by these cells, there would have to be at least two common numbers shared
by other cells.

Due to the definitions of $A_{ij}'$ and $B_{ij}'$
and the fact that the columns of $B_{ij}$ contain all distinct numbers $0,...,d-1$,
every row of the table contains $d^2$ distinct numbers $0,...,d^2-1$.
By construction, the numbers of any cell of the coordinate row
are distributed among all the cells of all the other rows.
Therefore, it is sufficient to prove the property of the net for the remaining rows.
Assume to the contrary, that two numbers repeat in two cells of different rows,
$(jd + B_{ij},j' d + B_{ij'}) = (l d + C_{kl}, l' d + C_{kl'})$.
Since $j,j',l,l',B_{ij},C_{ij} \in \{0,...,d-1\}$ the equality can only hold if $B_{ij} = C_{kj}$ and $B_{ij'} = C_{kj'}$,
i.e., there are rows of the squares $B$ and $C$ which contain the same numbers,
in the columns defined by $j$ and $j'$.
This, however, cannot be because one can always permute
the entries of, say, square $C$ such that its $k$th row
becomes the $i$th row (without compromising orthogonality)
and the two squares would not be orthogonal.

\section{Qubit}

Consider the squares for $d=2$.
We link them with the complementary measurements of a qubit.
The augmented set of orthogonal squares reads as
\begin{equation}
\begin{array}{cc}
0 & 1 \\
0 & 1
\end{array}
\qquad
\begin{array}{cc}
0 & 0 \\
1 & 1
\end{array}
\qquad
\begin{array}{cc}
0 & 1 \\
1 & 0.
\end{array}
\end{equation}
The right-hand side square is Latin, the left and middle square are orthogonal to each other
and to the Latin square.
These three squares lead to the following net design on the left-hand side,
in which the numbers are represented by pairs $mn$
in modulo-two decomposition:
\begin{equation}
\begin{tabular}[b]{cc|cc}
\multicolumn{2}{c|}{$b=0$} & \multicolumn{2}{c}{$b=1$} \\
\hline  \hline
$00$ & $01$ & $10$ & $11$
\\ \hline 
$00$ & $10$ & $01$ & $11$
\\ \hline 
$00$ & $11$ & $01$ & $10$
\\ \hline \hline
\end{tabular}
 \quad
\begin{tabular}[b]{c}
\hline  \hline
$m = b?$
\\ \hline 
$n = b?$
\\ \hline 
$m + n = b?$
\\ \hline \hline
\end{tabular}
\label{2DESIGN}
\end{equation}
On the right-hand side, we write down the
complementary questions associated with each row.
They are answered by pairs $mn$ in the left- and right-hand column
of the net design (left column $\to$ answer $0$, right column $\to$ answer $1$).
In this way, the questions are linked to the orthogonal squares.

The complementary questions can be answered in quantum experiments involving MUBs.
Consider a device encoding parameters $m$ and $n$
via application of the unitary $\hat U = \hat \sigma_x^m \hat \sigma_z^n$.
When it acts on $| z \pm \rangle$ states,
they get a phase dependent on $n$
and are flipped $m$ times.
Thus, knowing the initial state,
a final measurement in the $\hat \sigma_z$ eigenbasis reveals $m$,
giving the answer to the first complementary question.
Similarly, taking $| x \pm \rangle$ and $| y \pm \rangle$ as initial states,
the results of $\sigma_x$ and $\sigma_y$ measurement answer
the second and the third complementary question, respectively.

\section{Prime dimensions}

For prime $d$ the net has $d+1$ rows.
The entries of the rows corresponding to the OLSs are generated
from the following formula:
\begin{equation}
n = a m + b,
\label{PRIME_QUESTION}
\end{equation}
where the integer $a = 1, ... ,d-1$ enumerates the rows of the table,
while the integer $b = 0,...,d-1$ enumerates different columns,
and the sum is modulo $d$. 
Additional two rows correspond to the questions about $m$ and $n$, respectively.
The table for the rows corresponding to the OLSs is built in the following way:

(i) Choose a row, $a$, and the column, $b$.

(ii) Vary $m = 0,...,d-1$ and compute $n$ using (\ref{PRIME_QUESTION}).

(iii) Write pairs $mn$ in the cell.

For example, for $d=3$, one has
\begin{equation}
\begin{tabular}[b]{ccc|ccc|ccc}
\multicolumn{3}{c|}{$b=0$} & \multicolumn{3}{c|}{$b=1$} & \multicolumn{3}{c}{$b=2$} \\
\hline  \hline
$00$ & $01$ & $02$
& $10$& $11$& $12$
& $20$& $21$& $22$
\\ \hline 
$00$ & $10$ & $20$ 
& $01$& $11$& $21$
& $02$& $12$& $22$
\\ \hline 
$00$ & $11$ & $22$ 
& $01$& $12$& $20$
& $02$& $10$& $21$
\\ \hline 
$00$ & $12$ & $21$ 
& $01$& $10$& $22$
& $02$& $11$& $20$
\\ \hline \hline
\end{tabular}
 \quad
\begin{tabular}[b]{c}
\hline  \hline
$m = b?$
\\ \hline 
$n = b?$
\\ \hline 
$n = m + b?$
\\ \hline 
$n = 2 m + b?$
\\ \hline \hline
\end{tabular}
\label{3DESIGN}
\end{equation}
The complementary questions are given on the right-hand side.
Different values of $b$ enumerate possible answers.

We shall see, again, that the complementary questions can be answered using MUBs.
Consider encoding of parameters $m$ and $n$
via application of $\hat U = \hat X^m \hat Z^n$,
where the Weyl-Schwinger operators $\hat X^m \hat Z^n$ span a unitary operator basis.
In the basis of $\hat Z$, denoted as $| \kappa \rangle$, the two elementary operators satisfy
\begin{equation}
\hat Z | \kappa \rangle = \eta_d^{\kappa} | \kappa \rangle, \qquad
\hat X | \kappa \rangle = | \kappa + 1 \rangle,
\end{equation}
where $\eta_d = \exp{(i 2 \pi/d)}$ is a complex $d$th root of unity.
For the same reasons as for a qubit, the first two
questions are answered by applying $\hat U$ on an eigenstates of $\hat Z$ and $\hat X$ operators,
and then by measuring the emerging state in these bases.

In all other cases the action of the device
is $\hat U = \hat X^m \hat Z^{a m + b} = \hat X^m \hat Z^{am} \hat Z^b$.
The elementary operators do not commute,
instead one has $\hat Z \hat X = \eta_d \hat X \hat Z$,
and it follows that $\hat X^m \hat Z^{am} = \eta_d^{-\frac{1}{2}am(m-1)} (\hat X \hat Z^a)^m$.
Finally, the action of the device is, up to the global phase, given by
$\hat U \propto (\hat X \hat Z^a)^m \hat Z^b$.
The eigenstates of  the $\hat X \hat Z^a$ operator, expressed
in the $\hat Z$ basis, are given by
$| j \rangle_a = (1/\sqrt{d}) \sum_{\kappa=0}^{d-1} \eta_d^{-j \kappa - a s_{\kappa}} | \kappa \rangle$,
where $s_{\kappa} = \kappa+...+(d-1)$ \cite{UNITARY_MUB},
and the $\hat Z$ operator shifts them, $\hat Z | j \rangle_a = | j - 1 \rangle_a$.
After the device, $| j \rangle_a$ is shifted exactly $b$ times
and subsequent measurement in this basis reveals the answer to the $a$th question.
On the other hand, the eigenbases of $\hat X \hat Z^a$ for $a=1,...,d-1$ and eigenbases of $\hat X$ and $\hat Z$ 
are known to form a complete set of MUBs \cite{UNITARY_MUB}.
Not only the number of MUBs is the same as the number of OLSs,
but they are indexed by the same variable, $a$.
This allows to associate MUB to every OLS for prime $d$.

\section{Powers of primes}

If $d$ is a power of a prime, a complete set of OLSs is obtained using operations in the finite field of $d$ elements,
and one expects that a complete set of MUBs also follows from the existence of the field.
Indeed, explicit formulae for MUBs in terms of the field operations were presented in \cite{WOOTTERSFIELDS,DURT,KR}.
Here, we prove this result in a simple way related to \cite{WOOTTERS_PHASE_SPACE},
using the theorem of Bandyopadhyay \emph{et al.}\ \cite{UNITARY_MUB,GRASSL}:
If there is a set of orthogonal unitary matrices,
which can be partitioned into $M$ subsets of $d$ commuting operators,
then there are at least $M$ MUBs.
They are the joint eigenbases of the $d$ commuting operators.

To illustrate the idea, consider again prime $d$.
Take the orthogonal unitary operators $\hat S_{mn} = \hat X^m \hat Z^n$ 
with their powers $mn$ taken from the first column of the net.
The cell of the first and second row corresponds to the eigenbases of $\hat Z$ and $\hat X$, respectively,
whereas the other two rows are defined by $b = 0$,  i.e., $n = am$.
According to the commutation rule of the elementary operators $\hat X$ and $\hat Z$,
$\hat S_{mn}$ and $\hat S_{m'n'}$ commute if and only if $m n' - m' n = 0 \mod d$.
Thus, for a fixed row, i.e., fixed $a$, the set of $d$ operators $\hat S_{mn}$ commute, because $m (a m') - m' (a m) = 0$,
and, due to the mentioned theorem, there is a set of $d+1$ MUBs.

For $d = p^r$ being a power of a prime,
the OLSs and the net are generated  by the formula
\begin{equation}
n = a \odot m \oplus b,
\label{POWER_QUESTION}
\end{equation}
where $\odot$ and $\oplus$ denote multiplication and addition in the field,
$a,b,m,n\in \mathbb{F}_d$ are field elements, and $a \ne 0$.
The first two rows of the table are defined by $m = b$ and $n = b$.
In the case of $d=4$, the four elements $\{0,1,\omega,\omega+1\}$ of the field $\mathbb{F}_4$ 
($\omega$ is the root of $x^2+x+1$ \cite{WOOTTERS_PHASE_SPACE}),
when indexed with the numbers $\{0,1,2,3\}$,
lead to the following net design:
\begin{equation}
\begin{tabular}{cccc|cccc|cccc|cccc}
\hline  \hline
$00$ & $01$ & $02$ & $03$
& $10$& $11$& $12$& $13$
& $20$& $21$& $22$& $23$
& $30$& $31$& $32$& $33$
\\ \hline 
$00$ & $10$ & $20$ & $30$
& $01$& $11$& $21$& $31$
& $02$& $12$& $22$& $32$
& $03$& $13$& $23$& $33$
\\ \hline 
$00$ & $11$ & $22$ & $33$
& $01$& $10$& $23$& $32$
& $02$& $13$& $20$& $31$
& $03$& $12$& $21$& $30$
\\ \hline 
$00$ & $12$ & $23$ & $31$
& $01$& $13$& $22$& $30$
& $02$& $10$& $21$& $33$
& $03$& $11$& $20$& $32$
\\ \hline 
$00$ & $13$ & $21$ & $32$
& $01$& $12$& $20$& $33$
& $02$& $11$& $23$& $30$
& $03$& $10$& $22$& $31$
\\ \hline \hline
\end{tabular}
\label{4DESIGN}
\end{equation}

We use the concept of a basis in the finite field $\mathbb{F}_d$.
It consists of $r$ elements $e_i$, with $i=1,...,r$.
Every basis has a unique dual basis, $\overline{e}_j$,
such that $\mathrm{tr}(e_i \odot \overline{e}_j) = \delta_{ij}$,
where the trace in the field, $\mathrm{tr}(x)$, maps elements of $\mathbb{F}_d$
into the elements of the prime field $\mathbb{F}_p$.
We use lowercase tr for the trace in the field
in order to distinguish it from the usual trace over an operator,
which we denote by Tr.
It has the following useful properties:
$\mathrm{tr}(x \oplus y) = \mathrm{tr}(x) + \mathrm{tr}(y)$,
and $\mathrm{tr}(a \odot x) = a \, \mathrm{tr}(x)$,
where operations on the right-hand side are modulo $p$
and $a$ is in the prime field.
We decompose $m$ in the basis $e_i$,
$m = m_1 \odot e_1 \oplus ... \oplus m_r \odot e_r$, where $m_i = \mathrm{tr}(m \odot \overline{e}_i)$,
and $n$ in the dual basis,
$n = n_1 \odot \overline{e}_1 \oplus ... \oplus n_r \odot \overline{e}_r$, with $n_i = \mathrm{tr}(n \odot e_i)$.
Due to the properties of the trace in the field and the dual basis
\begin{equation}
\mathrm{tr}(m \odot n) = \sum_{i=1}^r m_i n_i = \vec m \cdot \vec n,
\label{FIELD_TRACE}
\end{equation}
where $\vec m = (m_1,...,m_r)$ and $\vec n = (n_1,...,n_r)$
have components in the prime field, i.e., numbers $\{0,...,p-1\}$.

Consider operators defined by the decomposition of $m$ and $n$,
$\hat S_{\vec m \vec n} = \hat X_p^{m_1} \hat Z_p^{n_1} \otimes ... \otimes \hat X_p^{m_r} \hat Z_p^{n_r}$,
where, e.g.,\ $\hat X_{p}^{m_i}$ is the unitary operator acting on the $i$th $p$-dimensional subspace of the global $d$-dimensional space.
Operators $\hat S_{\vec m \vec n}$ form an orthogonal basis.
They commute, if and only if $\vec m \cdot \vec n' - \vec m' \cdot \vec n = 0 \mod p$.
Take the operators corresponding to a fixed row of the first column of the net,
i.e.,\ $a$ is fixed, $b = 0$ and therefore $n = a \odot m$.
From Eq. (\ref{FIELD_TRACE}), all these $d$ operators commute if
$\mathrm{tr}(m \odot a \odot m') = \mathrm{tr}(m' \odot a \odot m)$,
which is satisfied due to 
associativity and commutativity of multiplication in the field.
Therefore, their eigenbases define MUBs.
Again, each row of the table is linked with the MUB.

To make an illustration, consider again the example of $d=4$.
Choose $(e_1,e_2) = (\omega ,1)$ as a basis in the field,
such that the numbers $m$ are decomposed into pairs $m \to m_1\, m_2$
in the usual way: $0 \to 0\,0$, $1 \to 0\,1$, $2 \to 1\,0$, $3 \to 1\,1$.
The dual basis reads as $(\overline{e}_1,\overline{e}_2) = (1,\omega + 1)$,
which implies that the numbers $n$ are decomposed into pairs $n \to n_1\, n_2$ as follows:
$0 \to 0\,0$, $1 \to 1\,0$, $2 \to 1\,1$, $3 \to 0\,1$.
Each pair of numbers of table (\ref{4DESIGN}) is now written
vertically as a combination of two pairs of numbers:
\begin{equation}
\begin{tabular}{cccc|cccc|cccc|cccc}
\hline  \hline
$00$ & $01$ & $01$ & $00$
& $00$& $01$& $01$& $00$
& $10$& $11$& $11$& $10$
& $10$& $11$& $11$& $10$
 \\
$00$ & $00$ & $01$ & $01$
& $10$  & $10$  & $11$  & $11$
& $00$  & $00$  & $01$  & $01$
& $10$  & $10$  & $11$  & $11$
\\ \hline 
$00$ & $00$ & $10$ & $10$
& $01$& $01$& $11$& $11$
& $01$& $01$& $11$& $11$
& $00$& $00$& $10$& $10$
\\
$00$ & $10$ & $00$ & $10$
& $00$  & $10$ & $00$  & $10$
& $01$  & $11$  & $01$  & $11$
& $01$ & $11$  & $01$ & $11$
\\ \hline 
$00$ & $01$ & $11$ & $10$
& $01$& $00$& $10$& $11$
& $01$& $00$& $10$& $11$
& $00$& $01$& $11$& $10$
\\
$00$ & $10$ & $01$ & $11$
& $00$  & $10$  & $01$  & $11$
& $01$  & $11$  & $00$  & $10$
& $01$  & $11$  & $00$  & $10$
\\ \hline 
$00$ & $01$ & $10$ & $11$
& $01$& $00$& $11$& $10$
& $01$& $00$& $11$& $10$
& $00$& $01$& $10$& $11$
\\
$00$ & $11$ & $01$ & $10$
& $00$  & $11$  & $01$  & $10$
& $01$  & $10$  & $00$  & $11$
& $01$  & $10$  & $00$  & $11$
\\ \hline 
$00$ & $00$ & $11$ & $11$
& $01$& $01$& $10$& $10$
& $01$& $01$& $10$& $10$
& $00$& $00$& $11$& $11$
\\
$00$ & $11$ & $00$ & $11$
& $00$  & $11$  & $00$  & $11$
& $01$  & $10$  & $01$  & $10$
& $01$  & $10$  & $01$  & $10$
\\ \hline \hline
\end{tabular}
\end{equation}
MUBs are formed by the eigenbases of operators $\hat \sigma_x^{m_1} \hat \sigma_z^{n_1} \otimes \hat \sigma_x^{m_2} \hat \sigma_z^{n_2}$,
where the powers are taken from the first column of this table.
The result is in agreement with other methods \cite{CASLAVS,UNITARY_MUB}.
The complementary questions answered by the states of these MUBs
are formulated in terms of individual bits $m_1$, $m_2$, $n_1$, $n_2$,
which are encoded by
$\hat U = \hat \sigma_x^{m_1} \hat \sigma_z^{n_1} \otimes \hat \sigma_x^{m_2} \hat \sigma_z^{n_2}$.
For example, the question of the last row is about the values of 
$m_1+ n_1$ and $m_2 + n_2$.

An interesting feature strengthening the link between MUBs and OLSs
is the existence of the set of OLSs and MUBs which cannot be completed.
For example, the following net design
\begin{equation}
\begin{tabular}{cccc|cccc|cccc|cccc}
\hline  \hline
$00$ & $01$ & $02$ & $03$
& $10$& $11$& $12$& $13$
& $20$& $21$& $22$& $23$
& $30$& $31$& $32$& $33$
\\ \hline 
$00$ & $10$ & $20$ & $30$
& $01$& $11$& $21$& $31$
& $02$& $12$& $22$& $32$
& $03$& $13$& $23$& $33$
\\ \hline 
$00$ & $11$ & $22$ & $33$
& $01$& $12$& $23$& $30$
& $02$& $13$& $20$& $31$
& $03$& $10$& $21$& $32$
\\ \hline \hline
\end{tabular}
\label{INCOMPLETE}
\end{equation}
cannot have more rows.
The MUBs related to this table
are the eigenbases of $\hat X$, $\hat Z$, and $\hat X \hat Z$ for $d=4$.
Correspondingly, there are no other bases which are mutually unbiased with respect to these three \cite{GRASSL_P}.

\section{General dimension}

Tarry was the first to prove that no two OLSs of order six exist \cite{TARRY},
i.e.,\ the net for $d=6$ has only three rows.
The operators $\hat X^m \hat Z^n$ commute for numbers $m$ and $n$ from the first cell of these rows
and the corresponding MUBs are the eigenbases of $\hat X$, $\hat Z$, and $\hat X \hat Z$.
Similarly to the case of $d=4$ no other MUB with respect to these three exists \cite{GRASSL}.
Of course, the question whether different three MUBs can be augmented
with additional MUBs remains open.

\subsection{MacNeish's bound}

More generally, the lower bound on the number of OLSs
was given by MacNeish \cite{MACNEISH}.
If two squares of order $a$ are orthogonal, $A \perp B$,
and two squares of order $b$ are orthogonal, $C \perp D$,
then the squares obtained by a direct product, of order $a b$, are also orthogonal, $A \times C \perp B \times D$.
This implies that the number of OLSs, $\mathcal{L}$,
of the order $d = p_1^{r_1}...p_n^{r_n}$, with $p_i$ being prime factors of $d$,
is at least $\mathcal{L} \ge \min_{i}(p_i^{r_i}-1)$, where $p_i^{r_i}-1$ is the number of OLSs of order $p_i^{r_i}$.
A parallel result holds for MUBs \cite{KR,GRASSL}.
If $| a \rangle$ and $| b \rangle$ are the states of two MUBs in dimension $d_1$,
and $| c \rangle$ and $| d \rangle$ are the states of MUBs in dimension $d_2$,
then the tensor product bases $| a \rangle \otimes | c \rangle$ and $| b \rangle \otimes |d \rangle$ 
form MUBs in dimension $d_1 d_2$.
Thus, for $d = p_1^{r_1}...p_n^{r_n}$ there are at least $\min_{i}(p_i^{r_i}+1)$ MUBs.

\subsection{Latin operator basis}

In general, we know more about the number of OLSs than about the number of MUBs \cite{DESIGN}.
We use this knowledge to derive conditions which restrict the form of MUBs.
Consider the operators
\begin{equation}
\hat B_{n_0...n_{d}} = \hat \openone + \sum_{m=0}^{d} \sum_{\xi=1}^{d-1} \eta_d^{n_m \xi} \hat S_m^{\xi},
\label{LATIN_BASIS}
\end{equation}
where $n_m = 0,...,d-1$ and $\hat S_m^{\xi} = \sum_{j=0}^{d-1} \eta_d^{j \xi} |j \rangle_m \langle j |$
have complete set of MUBs as eigenbases, $m=0,...,d$.
We show that existence of such a set
and orthogonality of $d^2$
operators $\hat B_{n_0...n_{d}}$ 
implies completeness of the set of OLSs.
The trace scalar product 
$\mathrm{Tr}(\hat B_{n_0...n_d}^{\dagger} \hat B_{n_0'...n_d'})$ is given by $d^2(k-1)$,
where $k$ denotes the sum of Kronecker deltas, $k \equiv \delta_{n_0 n_0'}+...+\delta_{n_d n_d'}$.
Operators $\hat B_{n_0...n_d}$ and $\hat B_{n_0'...n_d'}$ are orthogonal if and only if $k=1$,
i.e.\ $n_m = n_m'$ for exactly one $m$.
This condition applied to $d^2$ orthogonal operators,
defines a complete set of orthogonal squares.
To see this, take $d^2$ orthogonal operators $\hat B_{n_0(b)...n_d(b)}$ with $b=1,...,d^2$
and consider $d+1$ squares defined by their indices $n_m(b)$ for a fixed $m$.
If the squares were not orthogonal, one could find at least two identical pairs,
$(n_m(b),n_{m'}(b)) = (n_m(b'),n_{m'}(b'))$,
implying that operators (\ref{LATIN_BASIS}) are not orthogonal ($k >1$).
Therefore, e.g., for $d=6$, 
there is no complete set of MUBs
for which operators $\hat B_{n_0...n_d}$ are orthogonal
because 
there is no complete set of OLSs in this case.

\subsection{Orthogonal functions}

The second condition is obtained by noting that a net
defines ``orthogonal'' functions, $F_{a}(m,n)$,
which give the column of the $a$th row where the pair $m \, n$ is entered.
The orthogonality means that 
for the pairs $m\,n$ for which the function $F_{a}(m,n)$ has a fixed value, 
the function $F_{a'}(m,n)$ acquires all its values.
We show that if $d^2$ unitaries, $\hat U_{mn}$,
shift (up to a phase) the states of different bases in accordance with the net
\begin{equation}
\hat U_{mn} | j \rangle_a \propto |j + F_a(m,n) \rangle_a,
\label{SHIFTING_U}
\end{equation}
then these bases are MUBs.
For the proof, note that $\sum_{i'=0}^{d-1} |_{a}\langle i | i' \rangle_{a'}|^2 = 1$.
From orthogonality of the functions,
this sum can be written as $\sum_{\mathcal{S}} |_{a}\langle j + F_{a}(m,n) | j' + F_{a'}(m,n) \rangle_{a'}|^2=1$,
where $\mathcal{S}$ is the set of pairs $m \, n$ for which $F_{a}(m,n)$ has a fixed value.
By (\ref{SHIFTING_U}), the last is
$\sum_{\mathcal{S}} |_{a}\langle j | \hat U_{mn}^{\dagger} \hat U_{mn}| j' \rangle_{a'}|^2$,
which due to unitarity, $\hat U_{mn}^{\dagger} \hat U_{mn} = \openone$, is the sum of $d$
identical terms $|_{a}\langle j | j' \rangle_{a'}|^2$.
Therefore, $|_{a}\langle j | j' \rangle_{a'}|^2 = 1/d$.
Further, given $d^2$ unitaries with property (\ref{SHIFTING_U}),
one recovers the table in the following experiment:
Prepare $|0 \rangle_a$, act on it with $\hat U_{mn}$, measure in the same basis,
and write the pair $m \, n$ in the $a$th row and the column corresponding to the result.
Thus, in dimension six, there cannot be $36$ unitaries satisfying (\ref{SHIFTING_U}), with the orthogonal functions, for more than three bases,
because otherwise one could construct more than three orthogonal squares of order six, which is impossible.

\section{Hidden-variable simulation of MUBs}

The net designs can be used to construct hidden-variable models
which simulate results of complementary measurements on certain states.
Recently, Spekkens showed that only four ``ontic states'' (hidden variables)
are sufficient to simulate complementary measurements of a qubit prepared in a state of a MUB \cite{SPEKKENS}.
In his model, quantum states of MUBs correspond to the ``epistemic states'' satisfying the knowledge balance principle:
The amount of knowledge one possesses about the ontic state
is equal to the amount of knowledge one lacks \cite{SPEKKENS}.
This principle lies behind the net design.
Left table of (\ref{2DESIGN}) corresponds to the original Spekkens' model:
The numbers enumerate ontic states, cells correspond to the epistemic states
and rows to the complementary measurements.
All other tables generalize the model.
To identify the ontic state one needs two dits of information
(there are $d^2$ ontic states),
whereas the epistemic state is defined by a single dit,
leaving the other one unknown.
The quantum states described by these models require
(a classical mixture of) only two dits to model $d$ outcomes of $d+1$ quantum complementary measurements.

Our approach allows us to ask the question how many epistemic
states satisfying the knowledge balance principle,
i.e., having $d$ underlying ontic states, correspond to quantum states.
For example, in the case of a two-level system there are four ontic states,
and six possible epistemic states [see the net design of (\ref{2DESIGN})].
All six correspond to quantum eigenstates of complementary observables.
In general, any epistemic state is represented by a cell of $d$ numbers [$i_1 \ i_2 \ ... \ i_d$].
Since each number takes on one of $d^2$ values,
the numbers cannot repeat and their order is not important,
there are $\mathcal{E}_d = \sum_{i_1 = 1}^{D} \sum_{i_2 = i_1 + 1}^{D+1} \dots \sum_{i_d = i_{d-1} + 1}^{D+d-1}$ 
possible epistemic states, with $D = d^2 - d + 1$.

For $d$ being a power of a prime
the quantum states corresponding to the cells of the net design
are basis vectors of a complete set of MUBs.
They can be used to uniquely decompose arbitrary 
Hermitian operator
\begin{equation}
\hat O = - \mathrm{Tr}(\hat O) \hat \openone + \sum_{m=0}^d  \sum_{j=0}^{d-1} p_{j}^{(m)} |j \rangle_m \langle j|,
\label{MUB_EXPANSION}
\end{equation}
where $p_j^{(m)} = _m \!\! \langle j | \hat O | j \rangle_m$ and $|j \rangle_m$ is the $j$th state of the $m$th MUB.
For the proof, note that the complete set of MUBs can be used to define
the operator basis in the Hilbert-Schmidt space $\hat S_m^{\xi} = \sum_{j = 0}^{d-1} \eta_d^{j \xi} |j \rangle_m \langle j |$.
There are $d^2$ such operators, because $m=0,...,d$, the power $\xi=0,...,d-1$ 
and all $d$ operators $\hat S_m^0$ are equal to the identity operator.
Since they are normalized as $\mathrm{Tr}[(\hat S_m^{\xi})^{\dagger} \hat S_{m'}^{\xi'}] = d \delta_{m m'} \delta_{\xi \xi'}$
any operator has a unique expansion $\hat O = \frac{1}{d}[\mathrm{Tr}(\hat O) \hat \openone + \sum_{m=0}^d \sum_{\xi=1}^{d-1} \mathrm{Tr}(\hat O (\hat S_m^{\xi})^{\dagger}) \hat S_m^{\xi}]$
Writing $\hat S_m^{\xi}$ in terms of projectors on MUBs one finds Eq. (\ref{MUB_EXPANSION}).

If $\hat O$ is a quantum state, $\mathrm{Tr}(\hat O) = 1$ and $p_j^{(m)}$'s are probabilities to observe outcomes related to suitable states of MUBs.
We consider general epistemic states,
not necessarily those corresponding to the cells of the net design.
Such epistemic states have ``partial overlap'' with the cells,
defined as the number of common ontic states divided by $d$.
For example, the epistemic state [$00 \ 01 \ 20$]
has an overlap of $\frac{2}{3}$ and $\frac{1}{3}$
with the first and third epistemic state of the first row of table (\ref{3DESIGN}), respectively.
To construct operator $\hat O$ associated with a general epistemic state,
we take these overlaps to define the probabilities $p_j^{(m)}$.
Since we would like to see how many epistemic states correspond to quantum states
we take operators $\hat O$ with a unit trace.
If $\mathrm{Tr}(\hat O^2) = 1$ and $\mathrm{Tr}(\hat O^3) \ne 1$,
the operator $\hat O$ cannot represent a quantum state,
because the first condition excludes mixed states,
and both of them exclude pure states \cite{JL}.
We find that for $d=3$ only the epistemic states of the net design correspond to the quantum states.
There are $\mathcal{Q}_3 = 12$ such states, out of $\mathcal{E}_3 = 84$ different epistemic states.
The ratio of $\mathcal{R}_d = \mathcal{Q}_d / \mathcal{E}_d$ rapidly decreases with $d$:
we checked $\mathcal{R}_3 = 1/7$, $\mathcal{R}_4 = 8/455$ and $\mathcal{R}_5 = 1/1771$.
Thus, most of the epistemic states, constructed according to the ``knowledge balance principle,''
do not represent a quantum-physical state.

\section{Conclusions}

In conclusion,  we showed a one-to-one relation between OLSs and MUBs, if $d$ is a power of a prime.
For general dimensions, we derive conditions which limit the structure of the complete set of MUBs
and we presented parallelism between the MacNeish's bound on the minimal
number of OLSs and the minimal number of MUBs.
Interestingly, the MacNeish's bound is known not to be tight.
There are at least five OLSs of order $35$, where the MacNeish's bound is four \cite{WOJTAS}.
Therefore, further insight into the relations between MUBs and OLSs
would be gained from studies of MUBs for $d=35$.

Finally, using the squares, we constructed 
hidden-variable models that efficiently simulate measurements of MUBs.
However, the majority of states in these models do not have quantum-physical counterparts.

\section{Acknowledgments}

We thank Gabriele Uchida and Johannes Kofler.
This work is supported by the FWF Project No. P19570-N16
and project CoQuS (No. W1210-N16), and by
EC Project QAP (No. 015846),
and the Foundational Questions Institute (FQXi).

\end{document}